\documentstyle[12pt]{article}

\begin{document}

\baselineskip = 15pt

\def\la{\mathrel{\mathpalette\fun <}}
\def\ga{\mathrel{\mathpalette\fun >}}
\def\fun#1#2{\lower3.6pt\vbox{\baselineskip0pt\lineskip.9pt
  \ialign{$\mathsurround=0pt#1\hfil##\hfil$\crcr#2\crcr\sim\crcr}}}

\def\xvec{{\bf x}}
\def\kvec{{\bf k}}
\def\nvec{{\bf n}}
\def\qvec{{\bf q}}
\def\delt2{\langle \delta^2 \rangle}

\hskip 4in OSU-TA-22/95

\vskip 1.75in

\centerline{\bf A LINEAR PROGRAMMING APPROACH TO}

\centerline{\bf INHOMOGENEOUS PRIMORDIAL NUCLEOSYNTHESIS}

\vskip .2in

\centerline{Richard E. Leonard and Robert J. Scherrer}

\centerline{Department of Physics}

\centerline{The Ohio State University}

\centerline{Columbus, OH 43210}

\vskip 1in

\centerline{ABSTRACT}

We examine inhomogeneous primordial nucleosynthesis for {\it arbitrary}
distributions $f$ of the baryon-to-photon ratio $\eta$,
in
the limit where neither particle diffusion nor gravitational collapse
is important.
By discretizing $f(\eta)$ and using linear programming, we show
that for a set of $m$ constraints on the primordial element
abundances, the maximum and minimum possible values of $\bar \eta$ (the final
mean value of $\eta$)
are given when $f(\eta)$ is a sum of at most $m+1$ delta functions.
Our linear programming results indicate that when $f$ is taken to be an
arbitrary function,
there is no lower bound on $\bar \eta$, while the upper bound is essentially
the homogeneous upper bound.

\vskip 1 cm
\vfill
\eject

\centerline{1. INTRODUCTION}

\vskip 0.4 cm

One of the most widely-investigated variations in
primordial nucleosynthesis is the possibility that the baryon
density is inhomogeneous at the time of nucleosynthesis.  In the simplest
such inhomogeneous models, nucleosynthesis is assumed to take
place independently in regions of different density, with the final element
abundances derived by simply mixing the matter from these different
regions (Epstein \& Petrosian 1975; Barrow \& Morgan 1983; Yang et al. 1984;
Copi, Olive, \& Schramm 1995, hereafter COS).
Such a treatment is valid if the dominant fluctuations are of sufficiently
low amplitude and on a sufficiently large length scale that neither
gravitational collapse nor particle diffusion
is important.  If the length-scale of the fluctuations
is sufficiently short, then particle diffusion before and during
nucleosynthesis
can significantly alter the element yields (Applegate, Hogan, \& Scherrer 1987;
Alcock, Fuller, \& Mathews 1987; for more recent studies, see Thomas, et al.
1994;
Jedamzik, Fuller, \& Mathews 1994).  On the other hand, if the fluctuation
amplitudes are high enough, the highest density regions can undergo
gravitational collapse, effectively removing the elements in those regions from
the final measured abundances.  This scenario has also been extensively
explored (Sale \& Mathews 1986; Gnedin, Ostriker, \& Rees 1995; COS; Jedamzik
\& Fuller 1995).

Here we consider only the simplest of these scenarios (no diffusion or
collapse),
but we provide a significant generalization to previous work.  All previous
studies have assumed a particular form for the distribution
of baryon-to-photon ratio $f(\eta)$.  In this paper, we treat the case of
{\it arbitrary} $f(\eta)$.  Obviously, a brute-force calculation of
this case, examining all possible functions, would be impossible.
Even the analysis of a limited subset of the infinite number of possibilities
is an enormous computational task.
However, by discretizing the problem, we can turn it into a problem
in linear programming with a well-defined set of easily-obtained solutions.  In
the
next section, we explain our linear programming method and calculate the
density distributions which give the minimum and maximum possible values
for the mean value of $\eta$ for a reasonable set of observational constraints
on the primordial
element abundances.  Our conclusions are discussed in $\S$3.  We find
that there is no lower bound on $\eta$ (consistent with the earlier
arguments of Jedamzik, Mathews, \& Fuller 1995), while the upper bound
is essentially the homogeneous upper bound.

\vskip 0.6 cm
\centerline{2. CALCULATIONS}
\vskip 0.4 cm

The
most recent and complete treatment of inhomogeneous nucleosynthesis
when post-nucleosynthesis mixing is the only important effect
has been given
by COS, so we follow their treatment.
The elements produced in detectable quantities in primordial nucleosynthesis
are $^4$He, D, $^3$He, and $^7$Li.  The observational limits
on the abundances of these elements have
been discussed in great detail in a number of recent papers
(Walker et al. 1991; Copi, Schramm, \& Turner 1995a,b; Hata, et al. 1995,
1996).
However, since we wish to compare our results to the calculations of COS,
we will use the abundance limits given there:
\begin{eqnarray}
\label{abundance1}
0.221 &\le& {\rm Y_P} \le 0.243, \nonumber\\
({\rm D/H}) &\ge& 1.8 \times 10^{-5}, \nonumber\\
{\rm (D+ ^3He)/H} &\le& 1.0 \times 10^{-4}.
\end{eqnarray}
COS used two different bounds on the $^7$Li abundance:
\begin{eqnarray}
\label{abundance2}
({\rm ^7Li/H}) &\le& 1.4 \times 10^{-10}, \nonumber\\
({\rm ^7Li/H}) &\le& 2.0 \times 10^{-10},
\end{eqnarray}
Although we will use these limits in all of our calculations, our method
of calculation, which is the main new idea in this paper, can be applied to any
set
of limits.

Now assume that the distribution of densities is given by an unknown function
$f(\eta)$.
Previous studies have assumed a variety of functions for $f(\eta)$,
including the gamma distribution (Epstein \& Petrosian 1975; Yang et al. 1984)
and
the lognormal distribution (Barrow \& Morgan 1983).  COS considered both
of these as well as a Gaussian distribution for $\eta$.  We consider here the
case of arbitrary $f(\eta)$.  The mean final value for the baryon-to-photon
ratio, $\bar \eta$, is given by
\begin{equation}
\label{etabar}
\bar \eta = \int_0^\infty f(\eta) ~\eta ~d\eta
\end{equation}
The final element abundances are mass-weighted averages of the element
abundances produced in the individual regions.  If $\bar X_A$ is the mean
mass fraction of nuclide $A$ measured today, then
\begin{equation}
\label{Xbar}
\bar X_A = \int_0^\infty X_A(\eta) ~f(\eta)~ \eta~ d\eta/ \bar \eta,
\end{equation}
where $X_A(\eta)$ is the mass fraction of nuclide $A$
produced in standard (homogeneous) nucleosynthesis with
baryon-to-photon ratio $\eta$.  Since $f$ represents a probability
distribution, it must be normalized:
\begin{equation}
\label{normal}
\int_0^\infty f(\eta) ~d\eta = 1.
\end{equation}

Suppose that we wanted to test all possible functions $f$.  One way to do this
would be to divide the range in $\eta$ into discrete bins, and to approximate
the integrals in equations (\ref{etabar}) and (\ref{Xbar}) as sums:
\begin{equation}
\label{sumeta}
\bar \eta = \sum_i f_i \eta_i \Delta \eta_i
\end{equation}
and
\begin{equation}
\label{sumX}
\bar X_A = \sum_i {X_A}_i f_i \eta_i \Delta \eta_i/\bar \eta
\end{equation}
where the $\eta$ dependence of $f$ and $X_A$ is expressed through
their dependence on the bin number $i$.  The bins need not all
be of equal size, hence the factor $\Delta \eta_i$.
The normalization constraint (equation \ref{normal}) becomes
\begin{equation}
\label{sumnormal}
\sum_i f_i \Delta \eta_i = 1.
\end{equation}
Now one could imagine doing a Monte Carlo simulation, scanning through
all possible distributions $f_i$ which satisfy the constraint in equation
(\ref{sumnormal}).  In practice, this is impossible.  For example, if we
divided
the range in $\eta$ into 100 bins, and divided $f_i$ into 1000 ``units"
of magnitude 0.001
to be distributed among these bins, we would have to calculate
$C(1099,1000) \sim 10^{144}$ sets of element abundances, a calculation clearly
beyond all but the most robust graduate students.  However, most discussions
of inhomogeneous nucleosynthesis center on a much simpler problem:  given
a set of constraints such as those given in
equations (\ref{abundance1})-(\ref{abundance2}), and a function or
family of functions $f$ for the distribution of $\eta$, what are the largest
and smallest allowed values for $\bar \eta$? If we express the problem this
way, discretizing it as in equations (\ref{sumeta}) and (\ref{sumX}),
then the question of maximizing or minimizing the value of $\eta$ is
reduced to a problem in linear programming.  (A discussion of both
the theory and practice of linear programming can be found in, for example,
Press, Flannery, Teukolsky, \& Vetterling 1992, from which some
of the following discussion is taken).

The fundamental problem in linear programming is the following:
given a set of $N$ non-negative independent variables $x_j$,
and a set of $M$ constraints of the form:
\begin{equation}
\label{constraint1}
\sum_{j=1}^N a_j x_j \le b,
\end{equation}
or
\begin{equation}
\label{constraint2}
\sum_{j=1}^N a_j x_j \ge b,
\end{equation}
or
\begin{equation}
\label{constraint3}
\sum_{j=1}^N a_j x_j = b,
\end{equation}
maximize or minimize the function
\begin{equation}
z = \sum_{j=1}^N c_j x_j
\end{equation}
The linear programming nature of our discretized problem above becomes
clearer if we define the quantities $p_i$ to be
\begin{equation}
p_i \equiv f_i \Delta \eta_i.
\end{equation}
Then equations (\ref{sumeta})-(\ref{sumnormal}) become
\begin{equation}
\label{peta}
\bar \eta = \sum_i p_i \eta_i,
\end{equation}
\begin{equation}
\label{pX}
\bar X_A = \sum_i {X_A}_i p_i \eta_i /\bar \eta,
\end{equation}
and
\begin{equation}
\label{pnormal}
\sum_i p_i = 1.
\end{equation}
Equations (\ref{peta}) - (\ref{pnormal}) are now clearly in the form
of a linear programming problem:
the $N$ independent variables are the
$p_i$'s, equations (\ref{pX}) and (\ref{pnormal}) provide the constraint
equations, and $\bar \eta$ given by equation (\ref{peta}) is the quantity
we seek to maximize or minimize.  Note that equation (\ref{pX}) contains
$\bar \eta$ in the denominator, so that the bounds
$\bar X_A < X_{\rm upper bound}$ and $\bar X_A > X_{\rm lower bound}$ do not
immediately
translate into linear programming constraints of the form
given by equations (\ref{constraint1})-(\ref{constraint3}).
However, using
equations (\ref{peta}) and (\ref{pX}), we can rewrite these bounds
as
\begin{equation}
\sum_i [{X_A}_i - X_{\rm upper bound}] p_i \eta_i \le 0,
\end{equation}
and
\begin{equation}
\sum_i [{X_A}_i - X_{\rm lower bound}] p_i \eta_i \ge 0,
\end{equation}
which are in the form of equations (\ref{constraint1}) and (\ref{constraint2}).

An additional complication
is the fact that our constraints on all of the elements
other than $^4$He are expressed in terms of number ratios to hydrogen,
(A/H)$\equiv n_A/n_H$,
rather than as mass fractions, $X_A$, while our prescription for mixing
the various element abundances uses the mass fractions.  However, it is easy
to translate the number ratio bounds into a suitable form.  Recall
that
\begin{equation}
({\rm A/H}) = {X_A \over {\rm A} X_H},
\end{equation}
where $X_H$ is the $^1$H mass fraction.
Then an observational upper bound on (A/H) of the
form (A/H) $\le ({\rm A/H})_{\rm upper bound}$ can be written in the form
\begin{equation}
{\bar X_A \over A \bar X_H} \le ({\rm A/H})_{upperbound}
\end{equation}
where both $\bar X_A$ and $\bar X_H$ are given by equation (\ref{pX}).
Substituting for $\bar X_A$ and $\bar X_H$ from equation (\ref{pX}),
we obtain
\begin{equation}
\label{A/H1}
\sum_i [{X_A}_i - ({\rm A/H})_{\rm upper bound}{\rm A}{X_H}_i] p_i \eta_i
\le 0.
\end{equation}
Note that the expression for $\bar \eta$ has dropped out of the equation,
but now the inequality includes a sum over ${X_H}_i$.
For the lower bound (A/H) $\ge ({\rm A/H})_{\rm lower bound}$
we obtain a similar result:
\begin{equation}
\label{A/H2}
\sum_i [{X_A}_i - ({\rm A/H})_{\rm lower bound}{\rm A}{X_H}_i]p_i \eta_i \ge 0.
\end{equation}
Equations (\ref{A/H1}) and (\ref{A/H2}) are both in the form of acceptable
linear
programming constraints.

We now note an important result of linear programming theory:  given
a set of $N$ variables and $M$ constraints, the solution which maximizes
or minimizes $z$ has at least $N-M$ of the variables $x_j$ equal to zero
(Press et al. 1992).  In our case, if we have $m$ constraints
on the element abundances (e.g., equations (\ref{abundance1})
and (\ref{abundance2}) give $m=5$ constraints), and
the normalization condition (equation \ref{sumnormal}) provides
one additional constraint, then at most $M = m+1$ of the
$p_i$'s are non-zero; in this case $M=6$.
If we take the continuum limit of equations
(\ref{sumeta})-(\ref{sumnormal}), we arrive at the central
result of this paper:  for a set of $m$ constraints on the element
abundances such as those given in equations (\ref{abundance1})
and (\ref{abundance2}), and an arbitrary distribution of
$\eta$ given by the function $f(\eta)$, the largest
and smallest possible values for $\bar \eta$ occur when
$f(\eta)$ is the sum of at most $m+1$ delta functions.

We use the simplex method (Press et al. 1992) to solve our discretized
problem.  We ran the primordial nucleosynthesis
code of Wagoner (Wagoner, Fowler, \& Hoyle 1967) as updated
by Kawano (1992), with a neutron lifetime of $\tau = 887$ sec.
[This differs slightly from the neutron lifetime in COS, and we do
not include the small correction factor to the $^4$He production
used by COS, but these
are small differences which will not significantly affect our results].
We allowed $\eta$ to vary from $10^{-13}$ to $10^{-7}$, and
we divided this interval into 600 equally-spaced logarithmic
bins, calculating the mass fractions of $^4$He, $^3$He,
D, and $^7$Li for all 600 values of $\eta$.  We also included a bin
corresponding to zero baryon density and zero element production.
We then use the simplex
method to determine the form for $p_i$ which maximimizes or minimizes
$\bar \eta$ for the abundance constraints given in equations (\ref{abundance1})
and (\ref{abundance2}).

When we attempt to minimize $\bar \eta$, we find that almost all of $p_i$ is
concentrated in the lowest bin (i.e., the bin corresponding to zero baryon
density).
In fact, there is no ``lowest value" for $\bar \eta$;
the mean baryon-to-photon ratio can be arbitrarily small, a point recently
emphasized
by Jedamzik, Mathews, and Fuller (1995).  For example, one could take
$f(\eta) = p_1 \delta(\eta- \eta_0) + p_2 \delta (\eta)$, where $\eta_0$ is
a value for $\eta$ which gives acceptable nucleosynthesis for the homogeneous
case.
By mixing the correct homogeneous $\eta$ with the baryon-free regions, we
obtain
the correct element abundances regardless of the values of $p_1$ and $p_2$,
but by taking the limit $p_1 \rightarrow 0$, $p_2 \rightarrow 1$, the
value for $\bar \eta$ can be made arbitrarily small.

A more interesting question is the upper bound on $\bar \eta$.
Using either lithium bound in equation (\ref{abundance2}), we find
only two non-zero bins, straddling the upper bound on $\eta$ in the
homogeneous model.  To resolve this function further, we re-ran
the code using 1000 bins between $\eta = 10^{-10}$ and
$\eta = 10^{-9}$.
Using the first lithium bound in equation (\ref{abundance2})
we obtain:
\begin{equation}
f(\eta) = (0.21) \delta(\eta - 3.31 \times 10^{-10}) +
(0.79) \delta(\eta - 3.33 \times 10^{-10}),
\end{equation}
which gives $\bar \eta = 3.33 \times 10^{-10}$.
In this case, it is the limit on $^7$Li which is saturated.
The second lithium bound in equation (\ref{abundance2})
yields the solution:
\begin{equation}
f(\eta) = (0.58) \delta(\eta - 3.40 \times 10^{-10}) +
(0.42) \delta(\eta - 3.44 \times 10^{-10}),
\end{equation}
with $\bar \eta = 3.42 \times 10^{-10}$.  For this case, the $^4$He
limit is saturated.
In fact, given the precision with which we calculate
the various element abundances,
the difference between
these bounds and the homogeneous upper bound is not significant.
We are led to conclude that the homogeneous upper bound on $\eta$ cannot
be exceeded for ${\it any}$ distribution $f(\eta)$.
Our results are consistent with the claim that $f(\eta)$ should
be at most a sum of $m+1$ delta functions; in this case $m+1 = 6$, while
our optimum solution is the sum of only two delta functions.

Another interesting question is whether the inclusion of inhomogeneities can
resolve
a slight discepancy between the predictions of primordial nucleosynthesis and
the observations.
A number of recent studies (Copi, Schram, \& Turner 1995a,b; Hata, et al. 1995)
have
suggested that values of $\eta$ which give production of D, $^3$He, and $^7$Li
consistent
with the observations will overproduce $^4$He, so it has been suggested, for
example,
that the ``true" primordial $^4$He abundance is somewhat larger than is
currently
reported by the observers.  Since inhomogeneities allow for the reduction
of $\bar \eta$ by an arbitrary amount, is it tempting to think that
a reduction in the $^4$He abundance might also be possible.
To test this, we have used the D, $^3$He, and $^7$Li
limits given above, while minimizing the value of $^4$He.
Our linear programming method cannot be applied directly to minimize $^4$He,
because the expression for $^4$He contains a factor proportional to $\bar
\eta$.
However, we can find the
distribution in $p_i$ which minimizes
$^4$He for any given value of $\bar \eta$, so we have simply scanned over
a range in $\bar \eta$ to find the smallest $^4$He as a function of $\bar
\eta$.
Using this method, we find that no significant reduction in $^4$He is possible.
Again, our results apply to arbitrary distributions of $\eta$.

\vskip 0.6 cm

\centerline{3. DISCUSSION}

\vskip 0.4 cm

We find that for this simplest model for inhomogeneous nucleosynthesis, there
is no lower bound on $\eta$, as expected (Jedamzik, Mathews, \& Fuller 1995).
Previous studies which assumed particular functional forms for $f(\eta)$ all
produced a fairly narrow range in the allowed values for $\bar \eta$
(Epstein \& Petrosian 1975; Barrow \& Morgan 1983; Yang et al. 1984; COS).
However, this occurred because all of the functions $f(\eta)$ in these papers
were unimodal, i.e., characterized by a single maximum.
Hence, they cannot approximate the sort of solution which
has two large peaks at $\eta=0$ and $\eta= \eta_0$ (where $\eta_0$
gives acceptable abundances in the homogeneous model).
Our upper bound on $\bar \eta$ is, for all practical purposes, the homogeneous
upper bound on $\eta$.  This is consistent with the results of COS; for all
three functions they examined,
it is clear from their figures
that the upper bound on $\bar \eta$ is no larger than that obtained when the
variance of $f(\eta)$
goes to zero.
The importance of our results is that they give the most general upper and
lower
bounds on $\bar \eta$ for ${\it any}$ density distribution, essentially
bringing to a close the two-decade-long investigations of these simplest
inhomogeneous models.
Our final
conclusion is that for the inhomogeneous models in which
mixing is the only important process, $\bar \eta$ can be arbitrarily small
(a point already noted by Jedamzik, Mathews, \& Fuller 1995), but
$\bar \eta$ cannot be larger than the homogeneous upper bound on $\eta$.

Linear programming is not a technique usually applied to astrophysical
problems, although it has been previously used in galactic dynamics
(Schwarzschild 1979).
Our technique of discretizing
the problem and using linear programming could be applied to any
problem with constraints on integrals of an unknown function.
This technique cannot be applied to inhomogeneous
nucleosynthesis when particle diffusion is significant.  However,
it could be applied to the case when collapse of high density regions
is important.  We have chosen not to address this case because
it involves many particular model assumptions, but, for example,
equation (21) of COS can easily be put into the form of a linear programming
constraint.

\vskip 0.5in

We thank C. Copi, W. Press, and D. Thomas for helpful discussions.  R.E.L. and
R.J.S. were supported in part by the
Department of Energy (DE-AC02-76ER01545).  R.J.S was
supported in part
by NASA (NAG 5-2864).

\vfill
\eject
\centerline{\bf REFERENCES}

\vskip 1 cm

\noindent Alcock, C.R., Fuller, G.M., \& Mathews, G.J. 1987, ApJ, 320, 439

\noindent Applegate, J.H., Hogan, C.J., \& Scherrer, R.J. 1987, Phys Rev D,
35, 1151

\noindent Barrow, J.D., \& Morgan, J. 1983, MNRAS, 203, 393

\noindent Copi, C.J., Olive, K.A., \& Schramm, D.N. 1995, ApJ, 451, 51 (COS)

\noindent Copi, C.J., Schramm, D.N., \& Turner, M.S., 1995a, Science, 267, 192

\noindent Copi, C.J., Schramm, D.N., \& Turner, M.S., 1995b, Phys Rev Lett,
submitted

\noindent Epstein, R.I., \& Petrosian, V. 1975, ApJ, 197, 281

\noindent Gnedin, N.Y., Ostriker, J.P., \& Rees, M.J. 1995, ApJ, 438, 40

\noindent Hata, N., Scherrer, R.J., Steigman, G., Thomas, D., \& Walker, T.P.
1996, ApJ, in press

\noindent Hata, N., et al. 1995, Phys Rev Lett, in press

\noindent Jedamzik, K., \& Fuller, G.M. 1995, ApJ, 452, 33

\noindent Jedamzik, K., Fuller, G.M., \& Mathews, G.J. 1994, ApJ, 423, 50

\noindent Jedamzik, K., Mathews, G.J., \& Fuller, G.M. 1995, ApJ, 441, 465

\noindent Kawano, L., 1992, FERMILAB-Pub-92/04-A

\noindent Press, W.H., Flannery, B.P., Teukolsky, S.A., \& Vetterling, W.T.
1992, Numerical Recipes,

\indent  2nd edition, (Cambridge:  Cambridge University Press)

\noindent Sale, K.E., \& Mathews, G.J. 1986, ApJ, 309, L1

\noindent Schwarzschild, M. 1979, ApJ, 232, 236

\noindent Thomas, D., et al. 1994, ApJ, 430, 291

\noindent Wagoner, R.V., Fowler, W.A., and Hoyle, F. 1967, ApJ, 148, 3

\noindent Walker, T.P., Steigman, G., Schramm, D.N., Olive, K.A., \& Kang, H.
1991, ApJ, 376, 51

\noindent Yang, J., Turner, M.S., Steigman, G., Schramm, D.N., \& Olive,
K.A. 1984, ApJ 281, 493

\end{document}